\documentclass[twocolumn,final,aps,prb,showpacs,superscriptaddress,amsmath,amssymb,amsfonts,floatfix,longbibliography]{revtex4-1}
\usepackage{epsfig}
\usepackage[dvipsnames,usenames]{color}
\usepackage{color}
\usepackage{amsmath}
\usepackage{comment}
\usepackage{array}
\usepackage{multirow}
\usepackage{tabularx}
\usepackage{booktabs}
\usepackage{graphicx}
\usepackage{overpic}

\usepackage{float}
\usepackage{ulem}
\usepackage[english]{babel}
\usepackage[colorlinks=true,bookmarks=true, pdfborder={0 0 0},linkcolor=blue,urlcolor=blue,    breaklinks=true,bookmarksnumbered=true]{hyperref}
\usepackage{orcidlink}

\emergencystretch=\maxdimen
\begin{document}
\title{Role of interstitial $s$ orbital in a model of infinite-layer nickelates}
\title{Many-body calculation of band structure in a model of infinite-layer nickelates: (comparison with ARPES experiments) or \\
From 3D to quasi-2D: Strong-correlation-induced flattening of $k_z$ dispersion in infinite-layer nickelates }
\title{Role of interstitial $s$ orbital in a model of infinite-layer nickelates}
\author{Yan Peng \orcidlink{0009-0008-5604-7517}}
\affiliation{School of Physical Science and Technology, Soochow University, Suzhou 215006, China}
\author{Rui Peng}
\affiliation{State Key Laboratory of Surface Physics and Department of Physics, Fudan University, Shanghai 200433, P. R. China}
\author{Mi Jiang \orcidlink{0000-0002-9500-202X}}
\email[]{jiangmi@suda.edu.cn}
\affiliation{School of Physical Science and Technology, Soochow University, Suzhou 215006, China}
\affiliation{State Key Laboratory of Surface Physics and Department of Physics, Fudan University, Shanghai 200433, P. R. China}

\begin{abstract}
Motivated by recent angle-resolved photoemission spectroscopy (ARPES) experiments on infinite-layer (IL) nickelates, we employ determinant quantum Monte Carlo (DQMC) to study the three-orbital Emery model ($d$-$p$ model) coupled to an additional interstitial $s$ orbital retaining the three-dimensional dispersion. Our large-scale simulations reveal that: (1) the interstitial $s$-orbital-derived electron pocket is significantly reduced by the strong interaction but persists upon 20\% hole doping, reaching a size comparable to experimental observations; (2) the $d_{x^2-y^2}$-orbital dispersion is strongly renormalized by interactions, leading to a weak $k_z$ dependence consistent with ARPES measurements. 
Furthermore, compared with the conventional three-orbital $d$-$p$ model, the $d$-$p$-$s$ model exhibits enhanced short-range antiferromagnetic correlations.
These results highlight the crucial role of strong correlations and multi-orbital effects in shaping the low-energy electronic structure and many-body correlations in IL nickelates, and demonstrate the necessity of treating interaction-driven many-body physics within a realistic multi-orbital framework.
\end{abstract}

\maketitle

\section{Introduction}
Different from cuprates, in which the CuO$_2$ plane is widely believed to be the minimal motif in accounting for their low energy physics~\cite{anderson_resonating_1987, zhang_effective_1988, emery_theory_1987, lee_doping_2006}, the IL nickelates manifest more complex orbital structure~\cite{xie_microscopic_2022, gu_substantial_2020, liu_electronic_2020, yang_self-doping_2022, el-moudny_electronic_2023, sun_electronic_2025}, e.g. the rare-earth 5$d$ and 4$f$ orbitals. Hence, the extra physical phenomena, such as the Kondo effect~\cite{wang_kondo_2025, shao_kondo_2023}, Hund's coupling~\cite{kugler_low-energy_2024, kang_infinite-layer_2023}, self-doping on Ni-$d$ orbital~\cite{yang_self-doping_2022, di_cataldo_unconventional_2024} and the debate over its superconducting pairing symmetry~\cite{wang_distinct_2020, nomura_superconductivity_2022}, have sparked intense interest in the literature. Recent ARPES experiments have confirmed the existence of an interstitial $s$ orbital extending to the finite dopings in IL nickelates~\cite{ding_cuprate-like_2024, li_observation_2025}. On the contrary, no Ni-3$d_{z^2}$ character is detected on the Fermi surface. Furthermore, although a finite $k_z$ dispersion of the Ni $3d_{x^2-y^2}$ orbital is observed, its magnitude is significantly smaller than that predicted by density functional theory(DFT)~\cite{nomura_formation_2019, gu_substantial_2020, el-moudny_electronic_2023, kang_impact_2023}, and no Lifshitz transition is observed at the $Z$ point in $\mathrm{LaNiO_2}$, $\mathrm{NdNiO_2}$ and optimally doped $\mathrm{(La, Ca)NiO_2}$. The $\Gamma$-centered electron pocket with dominant $s$ orbital character does not emerge in ARPES results. Although some DFT studies~\cite{chen_electronic_2025, xia_three-dimensional_2025} have shifted the onsite energy of the $s$ orbital in order to remove this feature, the $d_{x^2-y^2}$ orbital still exhibits a pronounced $k_z$ dispersion that cannot be avoided. In fact, previous studies~\cite{jiang_stabilization_2022, PhysRevB.101.041104, liu_doping_2021} have supported the importance of multi-orbital physics in IL nickelate materials. In view of the limitations of these weak-coupling approaches, it is therefore essential to account for strong electronic correlations explicitly within the full many-body framework.

\begin{figure}[h!]
\centering
\psfig{figure=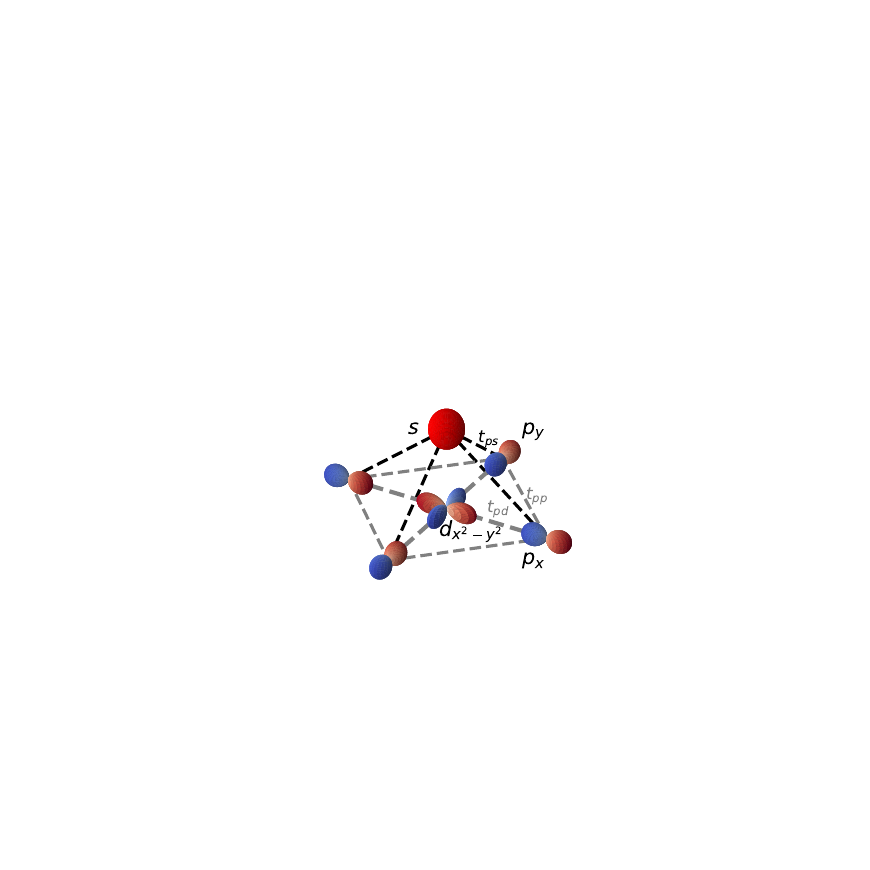,
width=0.99\columnwidth,
clip=,
trim=4.5cm 4.9cm 4.5cm 6.9cm}
\hfill
\psfig{figure=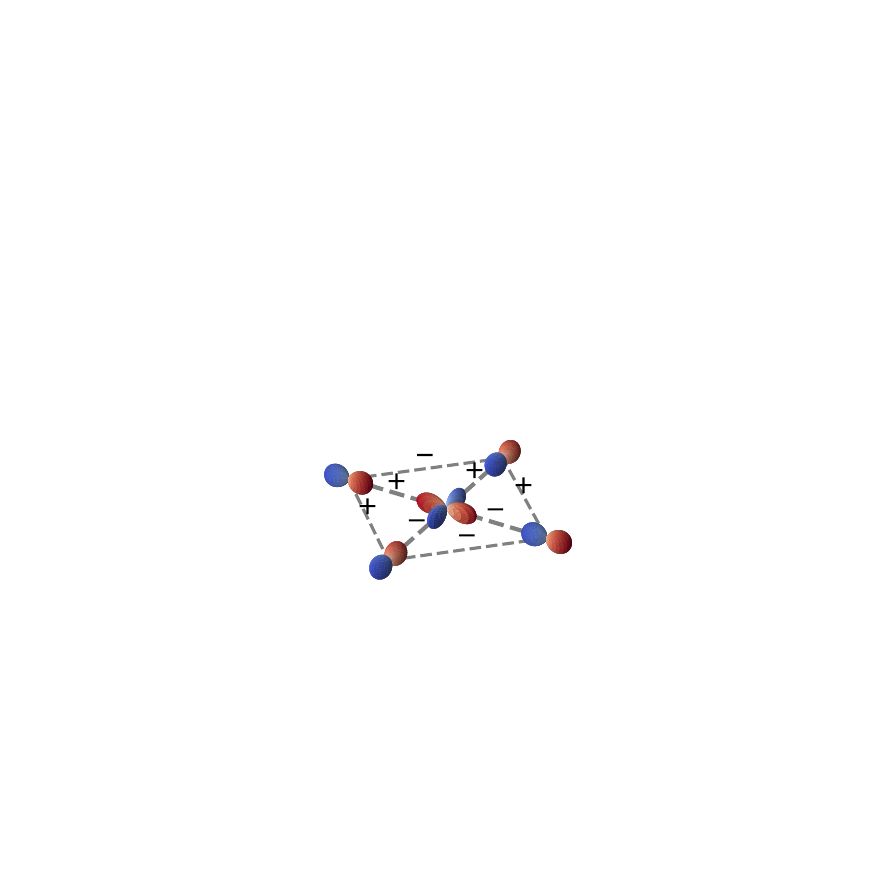,
width=0.99\columnwidth,
clip=,
trim=4.5cm 4.9cm 4.5cm 7.4cm}
\caption{Schematic illustration of (top) the $d$-$p$-$s$ model and (bottom) the $d$-$p$ (Emery) model. The signs of the corresponding orbital wavefunctions are indicated by red (positive) and blue (negative). Black and gray dashed lines distinguish the $d$-$p$ plane hybridization from the $p$-$s$ hybridization.}
\label{geom}
\end{figure}

As a numerically exact many-body method, DQMC has achieved remarkable success in accounting for the quasi-two-dimensional physics of cuprates for both the single-orbital Hubbard model~\cite{doi:10.7566/JPSJ.90.111010, wu_pseudogap_2018, qin_hubbard_2022} and the three-orbital Emery ($d$-$p$) model\cite{mai_fluctuating_2024, kung_characterizing_2016, huang_numerical_2017, peng_hole_2025, peng_interplay_2025}. Moreover, it remains applicable to relatively large lattice sizes at temperatures that are still rather low by numerical standards~\cite{huang_stripe_2018, mai_fluctuating_2024, peng_interplay_2025}. These capabilities enable us to examine the spectral properties at dominant $\mathbf{k}$ points in the Brillouin zone. In this work, based on the conventional Emery model, we explicitly add into the additional metallic interstitial $s$ orbital of IL nickelates by including its full three-dimensional dispersion. This partially mitigates the fermion sign problem because of the non-interacting nature of $s$ orbital, while at the same time preserving the essential underlying physics. Our simulations provide compelling evidence that strong electronic correlations substantially suppress the $k_z$-dispersion of the band structure. Simultaneously, the renormalization of the Fermi level reduces the electron occupation in the interstitial $s$ orbital but does not completely pushes its band below the Fermi level, bringing the calculated electronic structure into remarkable agreement with ARPES measurements. Enhanced short-range antiferromagnetic spin correlations are widely believed to promote unconventional superconductivity through spin fluctuation mediated pairing interactions. Consistent with this scenario, our results also show  the strengthening of short-range antiferromagnetic spin correlations.

The remainder of this paper is organized as follows. Section \ref{Model and Method} introduces the four-orbital $d$-$p$-$s$ model and DQMC method. Section \ref{sec:results} presents the numerical results and analysis, focusing on the orbital-resolved momentum-dependent spectral functions as well as the spin correlations. Finally, we conclude in Section \ref{sec:conclusion} with a summary and discussion.

\section{Model and Methodology}
\label{Model and Method}

\subsection{Hamiltonian and bare dispersion}

\begin{table}[t]
  \centering
  \caption{Parameters used in the models.}
  \label{tab:parameters}
  \begin{ruledtabular}
  \begin{tabular}{lcc}
    Parameter & Value & Description \\
    \hline

    $t_{pd}$ & 1.0 & NN $d$--$p$ hopping \\
    $t_{pp}$ & 0.4 & NN $p$--$p$ hopping \\
    $t_{ps}$ & 0.4 & NN $p$--$s$ hopping \\

    \hline

    $t_{ss}^{100}$ & 0.01 & $s$--$s$ hopping along $(100)$ \\
    $t_{ss}^{110}$ & 0.13 & $s$--$s$ hopping along $(110)$ \\
    $t_{ss}^{001}$ & 0.33 & $s$--$s$ hopping along $(001)$ \\
    $t_{ss}^{101}$ & 0.5 & $s$--$s$ hopping along $(101)$ \\
    $t_{ss}^{111}$ & -0.2 & $s$--$s$ hopping along $(111)$ \\

    \hline

    $U_{dd}$ & 6.0 & interaction on $d$ orbital \\
    $U_{pp}$ & 0 & interaction on $p$ orbital \\

    \hline

    $\epsilon_d$ & 0 & on-site energy on $d$ orbital \\
    $\epsilon_p$ & 5.0 & on-site energy on $p$ orbital \\
    $\epsilon_s$ & -3.0 & on-site energy on $s$ orbital \\

    \hline

    $N_c$ & up to $8\times8\times2$ & cluster size \\
    $\beta t_{pd}$ & up to 10.0 & inverse temperature \\

  \end{tabular}
  \end{ruledtabular}
\end{table}

We adopt the three-orbital $d$-$p$ model plus an extra interstitial $s$ orbital, which only hybridizes with nearest neighbor $p$ orbitals. This four-orbital Hamiltonian reads as
\begin{align}
H_0
&=
\sum_{\mathbf{k},\sigma}
\Psi_{\mathbf{k}\sigma}^\dagger
\,
\mathcal{H}(\mathbf{k})
\,
\Psi_{\mathbf{k}\sigma},
\end{align}
with
\begin{align}
\Psi_{\mathbf{k}\sigma}
&=
\begin{pmatrix}
d_{\mathbf{k}\sigma} \\
p_{x,\mathbf{k}\sigma} \\
p_{y,\mathbf{k}\sigma} \\
s_{\mathbf{k}\sigma}
\end{pmatrix}
\end{align}
and
\begin{align}
\mathcal{H}(\mathbf{k})=
\begin{pmatrix}
\epsilon_d-\mu
& -2it_{pd}s_x
& 2it_{pd}s_y
& 0
\\
2it_{pd}s_x c_z
& \epsilon_p-\mu
& 4t_{pp}s_x s_y
& 4it_{ps}s_x c_z
\\
-2it_{pd}s_y c_z
& 4t_{pp}s_x s_y
& \epsilon_p-\mu
& 4it_{ps}s_y c_z
\\
0
& -4it_{ps}s_x c_z
& -4it_{ps}s_y c_z
& \epsilon_s(\mathbf{k})-\mu
\end{pmatrix}
\end{align}
where
\begin{align}
s_x &= \sin\frac{k_x}{2},\qquad
s_y = \sin\frac{k_y}{2},\qquad
c_z = \cos\frac{k_z}{2}
\end{align}
and
\begin{align}
\epsilon_s(\mathbf{k})
= \epsilon_s
&+ 2 t_{ss}^{100} \left( \cos k_x + \cos k_y \right) \notag \\
&+ 4 t_{ss}^{110} \cos k_x \cos k_y \notag \\
&+ 2 t_{ss}^{001} \cos k_z \notag \\
&+ 4 t_{ss}^{101} \left( \cos k_x + \cos k_y \right) \cos k_z \notag \\
&+ 8 t_{ss}^{111} \cos k_x \cos k_y \cos k_z
\end{align}
Here $\epsilon_d$, $\epsilon_p$, and $\epsilon_s$ denote the onsite energies of the corresponding orbitals.
The hopping $t_{pd}$ describes  the hybridization between the Ni $d_{x^2-y^2}$ orbital and the surrounding oxygen $p$ orbitals; while $t_{pp}$ represents hopping between nearest neighboring oxygen sites within the NiO$_2$ plane. 
The interstitial $s$ orbital hybridizes with nearby oxygen orbitals through $t_{ps}$. The hybridization between the $d_{x^2-y2}$ and $s$ orbitals vanishes due to the symmetry constraint.
Fig.~\ref{geom} illustrates the geometry of our lattice and orbitals involved. The values of all model parameters relevant to the IL nickelates are summarized in Table~\ref{tab:parameters}. We emphasize that the $\epsilon_p=5.0$ is larger than the typical value relevant for cuprates to account for the larger charge transfer energy of infinite-layer nickelates~\cite{jiang_stabilization_2022,peng_interplay_2025}. 
Therefore, the doped holes are less likely to occupy the O orbitals, which lends support to our assumption of neglecting $U_{pp}$~\cite{mai_fluctuating_2024,kung_characterizing_2016, PhysRevX.10.021061}, which also practically mitigates the QMC sign problem.

\begin{figure}[t]
\centering
\begin{overpic}[width=0.99\columnwidth]{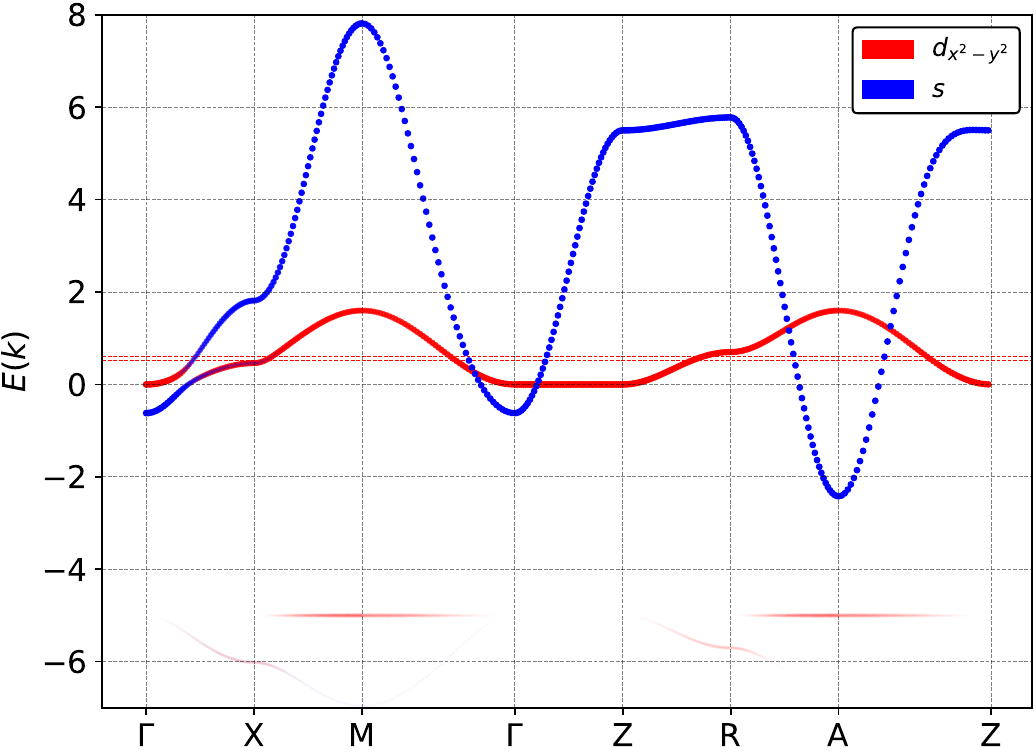}
\put(17,15){
\includegraphics[
width=0.3\columnwidth,
trim=2.0cm 0.cm 1.6cm 1.0cm,
clip
]{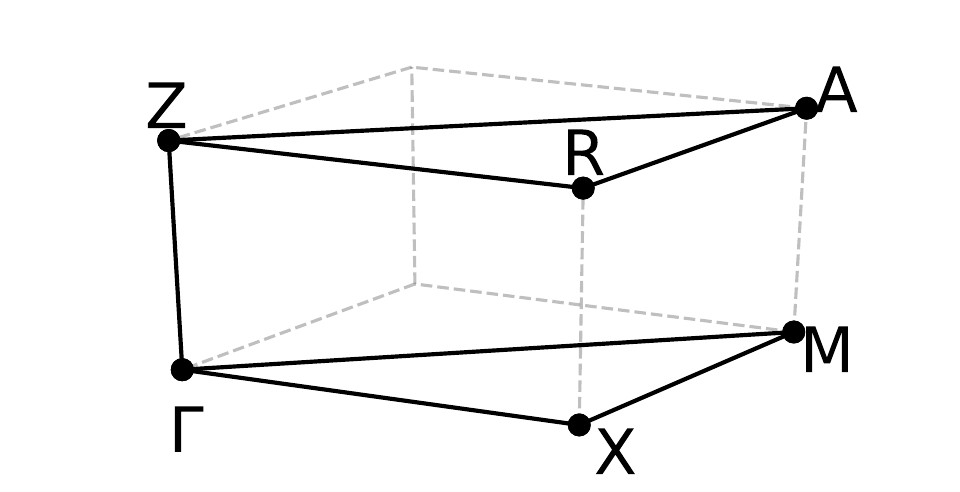}
}
\end{overpic}
\caption{The non-interacting band structure projected onto the $d_{x^2-y^2}$ and $s$ orbitals with the high-symmetry path displayed in the inset. Here the $d_{x^2-y^2}$ orbital exhibits a pronounced $k_z$ dispersion between points $X$ and $R$, while the interstitial $s$ orbital forms electron pockets at both the $\Gamma$ and $A$ points. The blurry bands between -5 and -6 eV originate from the hybridization to $p$ orbitals, whose dispersions are omitted since their energies lie far away from the Fermi level. The two red horizontal dashed lines indicate the Fermi levels corresponding to the parent compound LaNiO$_2$ (upper) and the 20\% Ca doped case (lower)~\cite{li_observation_2025}.}
\label{Fig2}
\end{figure}

Diagonalizing the Bloch Hamiltonian $\mathcal{H}(\mathbf{k})$ yields four non-interacting bands that serve as the starting point for the calculations incorporating the strong interaction. 
The non-interacting dispersions projected onto the relevant $d$ and $s$ orbitals are shown in Fig. \ref{Fig2}, whose inset shows the high-symmetry path of the Brillouin zone of the primitive cell.
Note that the $d_{x^2-y^2}$ orbital exhibits a pronounced $k_z$ dispersion between points $X$ and $R$; while the interstitial $s$ orbital forms electron pockets at both the $\Gamma$ and $A$ points. We choose the appropriate parameters in Table~\ref{tab:parameters} to host the above two features to be consistent with previous band structure calculations~\cite{el-moudny_electronic_2023, gu_substantial_2020, xia_three-dimensional_2025}. The two dashed lines indicate the Fermi levels corresponding to the parent compound LaNiO$_2$ and the 20\% Ca doped case in the ARPES experiments~\cite{ding_cuprate-like_2024, li_observation_2025}. As discussed later, this non-interacting band structure will be strongly renormalized by incorporating the strong interaction.

\subsection{DQMC and Maximum Entropy Method}

DQMC algorithm is a widely adopted many-body numerical technique to provide an unbiased evaluation of various physical properties of a system at finite temperatures~\cite{huang_numerical_2017, kung_characterizing_2016, doi:10.7566/JPSJ.90.111010, mai_fluctuating_2024}. It has been extensively employed in strongly correlated models and particularly has provided enormous understanding of the cuprate related physics. 
One of the main advantages of DQMC is that it treats the electron-electron interaction non-perturbatively and does not rely on uncontrolled approximations, making it a numerically exact approach apart from statistical error and finite-size effects. Within this framework, various thermodynamic and correlation functions can be computed directly, including single-particle Green's functions, spin and charge correlations, and pairing susceptibilities.
However, the notorious QMC sign problem limits the accessible parameter space so that most of our presented results would be for the inverse temperature $\beta t_{pd}=10.0$ as shown in Table~\ref{tab:parameters}. 

DQMC simulations evaluate both single- and two-particle Green's function (correlation functions) resolved in space and imaginary time~\cite{chang_recent_2015, blankenbecler_monte_1981, he_finite-temperature_2019}. 
The orbital-resolved local density of states (LDOS) can be obtained by analytically continuing the imaginary-time single-particle Green's function using Maximum Entropy Method (MEM)~\cite{gubernatis_quantum_1991, harremoes_maximum_2001, bergeron_algorithms_2016, wu_maximum_2012}
\begin{align}
    G_\alpha(\tau)=\int_{-\infty}^\infty d\omega\frac{e^{-\tau\omega}}{1+e^{-\beta\omega}} A_{\alpha}(\omega),
\end{align}
where $A_{\alpha}(\omega)$ represents the LDOS of orbital $\alpha$. Similarly, the momentum-resolved spectral function $A_\alpha(\mathbf{k},\omega)$ characterizes the dispersion of the correlated electronic states in momentum space of the corresponding orbital, which is comparable to the realistic ARPES experiments~\cite{li_observation_2025}.

\section{Results}
\label{sec:results}

Before proceeding, we remark that the following results adopt the hole language for $d$ and $p$ orbitals while electron language for the interstitial $s$ orbital. In particular, $n^h_{dp}=1.0$ corresponds to the half-filling situation as the conventional $d$-$p$ model; while $n^h_{dp}>1.0$ is for hole-doping. In addition, $n^e_s\sim0.05$ is adjusted to be quite small as revealed in the recent ARPES experiments~\cite{ding_cuprate-like_2024, li_observation_2025}.

For direct comparison to the experimental band structure, we focus on two characteristic combinations of the orbital occupancy, namely $n^h_{dp}\sim1.1, n^e_s\sim 0.05$ and $n^h_{dp}\sim1.25, n^e_s\sim 0.05$, to roughly correspond to the undoped and 20\% hole doped situations in experiments, respectively. 

To compromise the computational complexity (also limited by the QMC sign problem) and the inclusion of $k_z$ dependence of the renormalized band structure, we employ various lattice size, e.g $8\times8\times2$, $4\times4\times4$ for reliable calculations for band structures and smaller $4\times4\times2$ for more time-consuming spin correlation functions.

\begin{figure}[t]
\centering
\psfig{figure=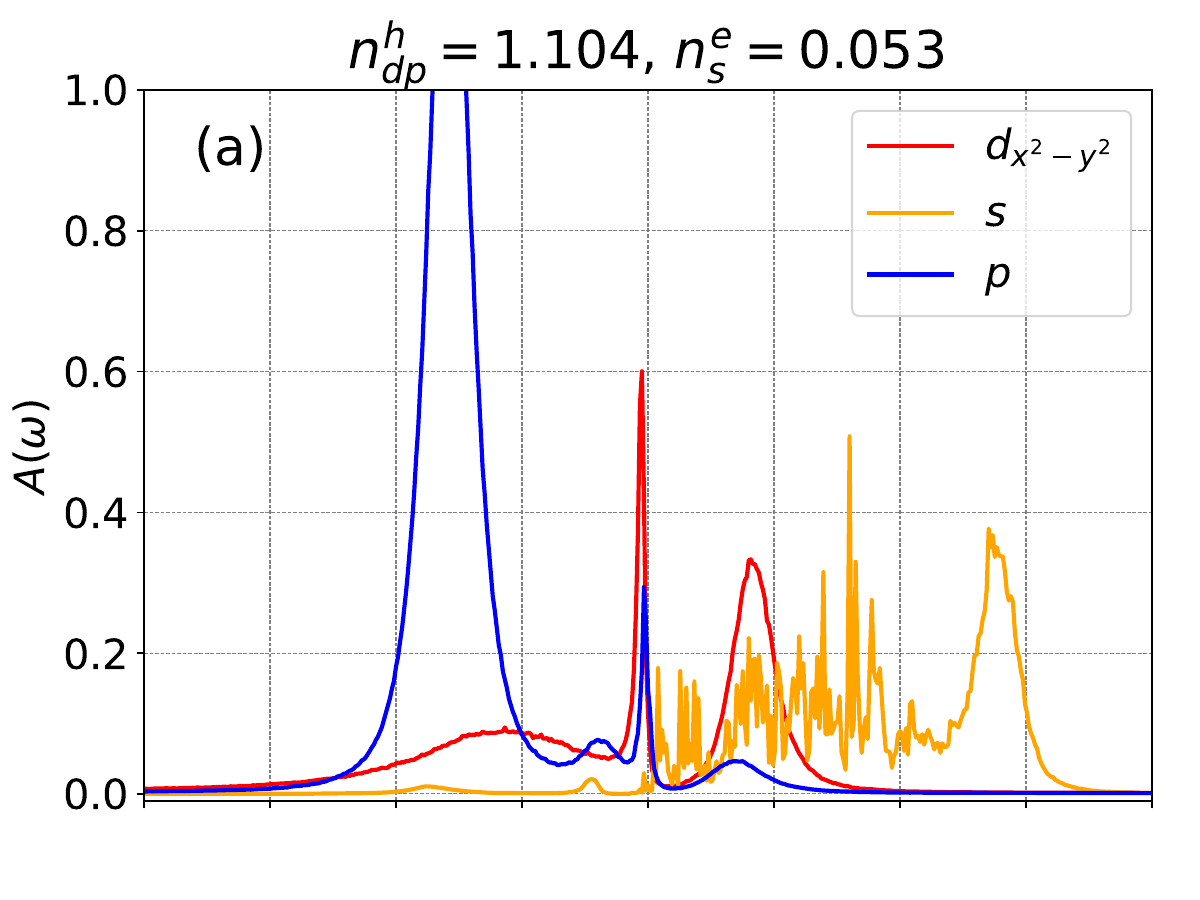, width=0.99\columnwidth, clip=,
trim=0cm 1.5cm 0cm 0cm}
\hfill
\psfig{figure=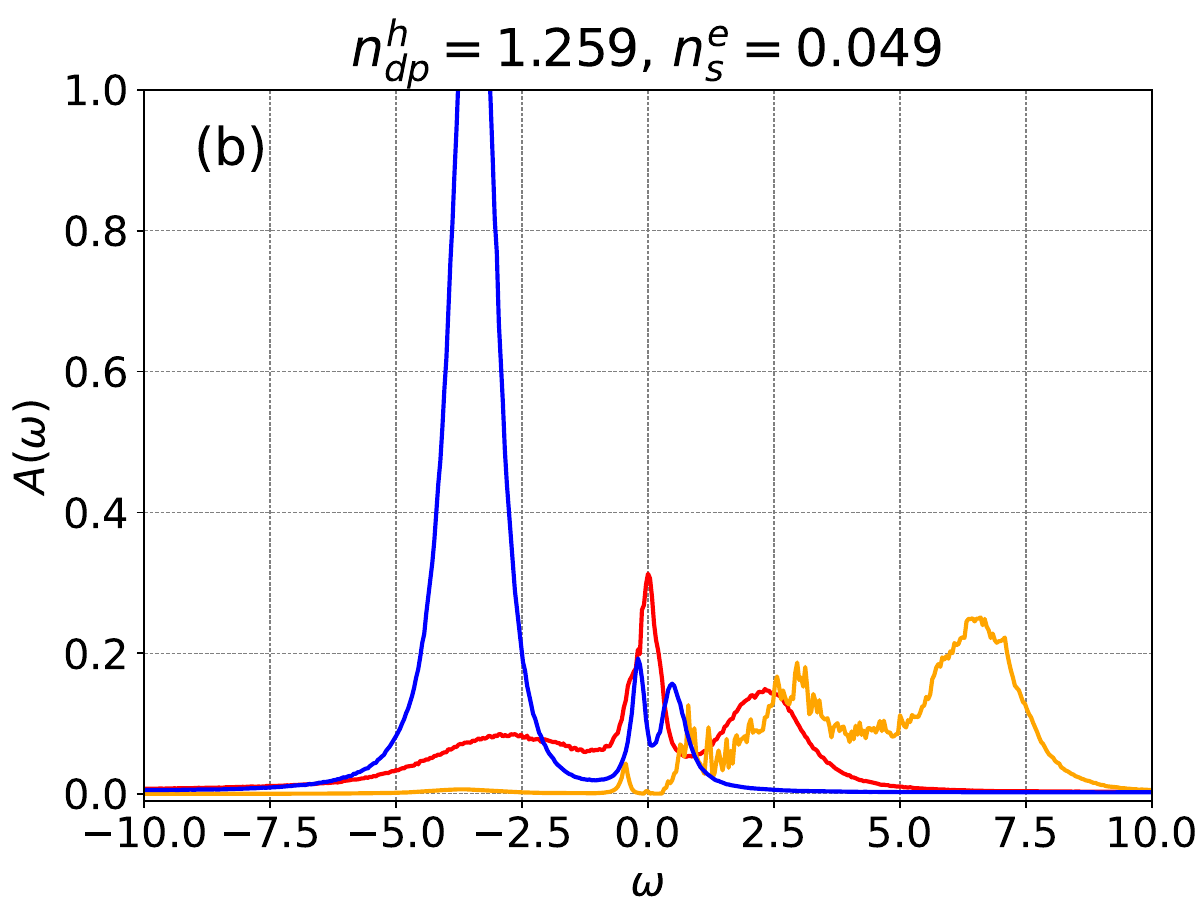, width=0.99\columnwidth}
\caption{Local density of states ($A(\omega)$) of the undoped and 20\% hole doped case. The hole density $n_{dp}^h$ in the $d$-$p$ plane and the electron density $n_s^e$ of the $s$ orbital are indicated above each panel.}
\label{Fig3}
\end{figure}

\subsection{Renormalized $s$ band}

Fig.~\ref{Fig3} compares the orbital-resolved LDOS for the undoped and 20\% hole doped cases. Apart from the broad O-$2p$ band around $-4\,\mathrm{eV}$, we focus on the features near the Fermi level.
For both cases, the small $n^e_s\sim0.05$ is reflected as the little peak around $-1\,\mathrm{eV}$ in the spectrum of $s$-orbital. Given that the local $n^e_s$ is already small, its electron density associated with the $A$ electron pocket of the Fermi surface becomes even smaller (compared to the large $A$ pocket originating from $s$ orbital in DFT's band structure of Fig.~\ref{Fig2}), bringing the theoretical estimate closer to the experimental observation~\cite{ding_cuprate-like_2024, li_observation_2025}. 

Compared with previous DQMC simulations for cuprate relevant parameters~\cite{kung_characterizing_2016}, the quasiparticle peak of the $d$ orbital at the Fermi level becomes noticeably sharper. This reflects the more localized nature of the $d_{x^2-y^2}$ orbital in IL nickelates. Additionally, upon doping, as shown in Fig.~\ref{Fig3}(b), the upper Hubbard band (UHB) transfers spectral weight towards lower energies, while the $p$-orbital peak that hybridizes with the $d$ orbital in the UHB shifts to lower energies. A similar behavior with hole doping has also been observed for the cuprate parameters~\cite{kung_characterizing_2016, peng_interplay_2025}. Finally, upon doping, $n^e_s$ slightly decreases from 0.053 to 0.049. The little peak shifts from about $-1\,\mathrm{eV}$ to around $-0.5\,\mathrm{eV}$, matching the shrinking of the electron pocket and consistent with ARPES experiments.

\begin{figure}[t]
\centering
\psfig{figure=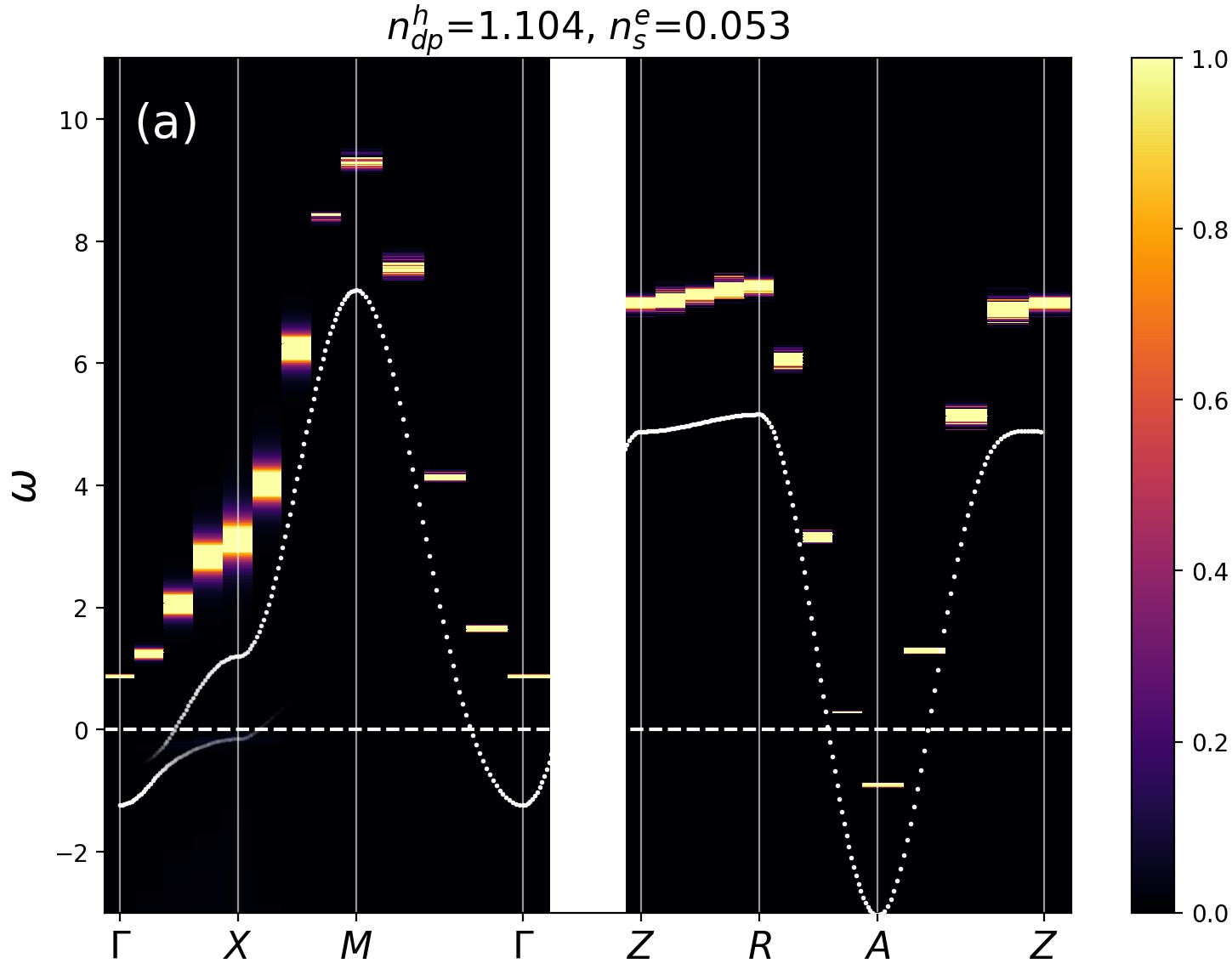, width=0.99\columnwidth}
\hfill
\psfig{figure=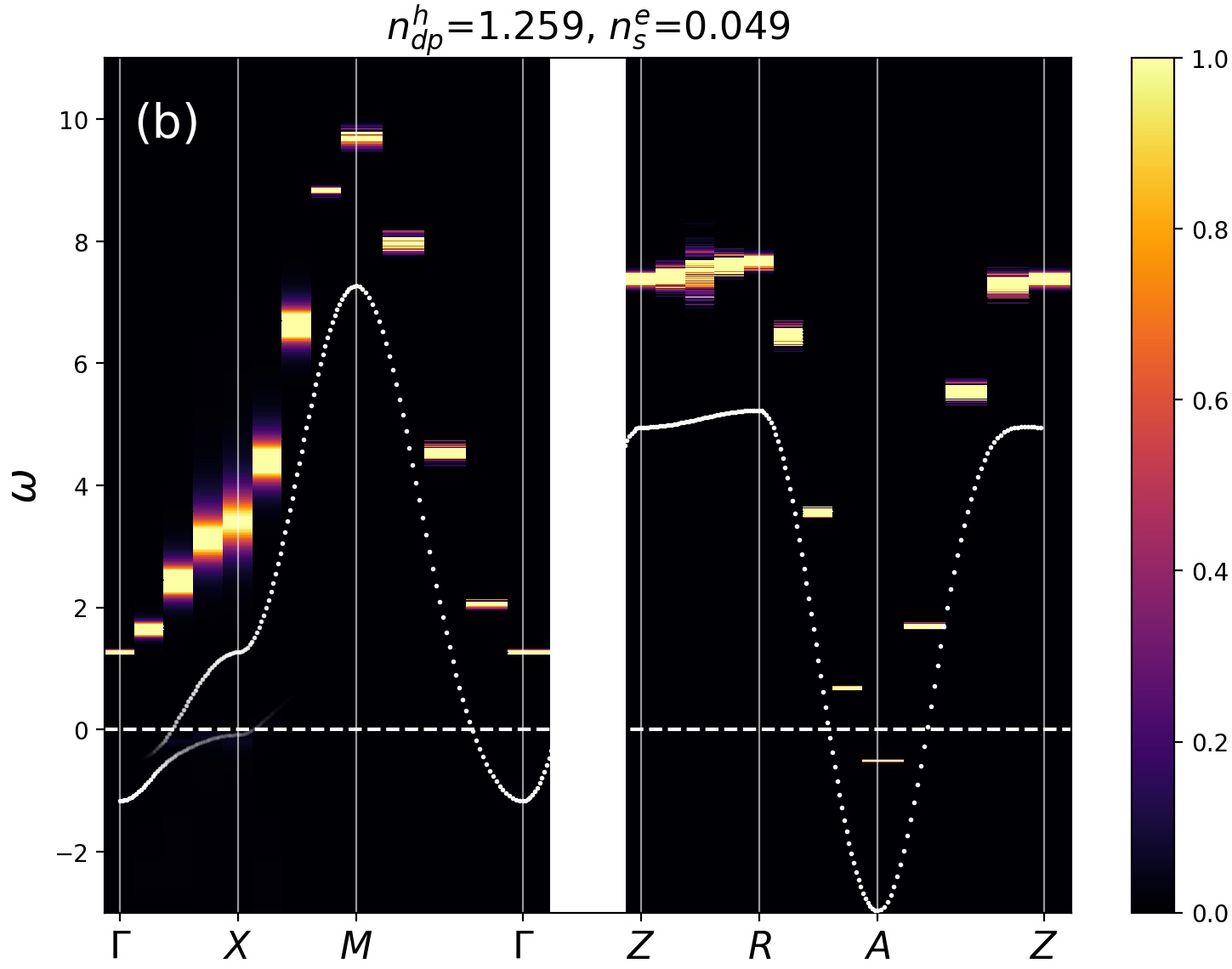, width=0.99\columnwidth}
\caption{Momentum-resolved spectral function $A_s(k,\omega)$ of the interstitial $s$ orbital along $\Gamma$-$X$-$M$-$\Gamma$ ($k_z=0$) and $Z$-$R$-$A$-$Z$ ($k_z=\pi$) for $8\times8\times2$ lattice. The hole density $n_{dp}^h$ in the $d$-$p$ plane and the electron density $n_s^e$ in the $s$ orbital are indicated above each panel. Panels (a) and (b) correspond to the parent compound and the 20\% hole-doped case in the ARPES experiment, respectively. The white dotted curves indicate the non-interacting dispersion.}
\label{Fig4}
\end{figure}

Fig.~\ref{Fig4} further shows the momentum-resolved spectra $A_s(\mathbf{k},\omega)$ of $s$ orbital along high symmetry path. It can be seen that, irrespective of doping, the electron pocket at the $\Gamma$ point is renormalized to energies above the Fermi level once strong correlations are taken into account, which coincides with the absence of features in ARPES detection. 

Another dominant feature is the significant shrink of $A$ electron pocket by the proper inclusion of strong correlations in our calculation. More importantly, the pocket only shrinks with hole doping but does not disappear even at 20\% hole doping, which is remarkably in agreement with the observation of ARPES~\cite{li_observation_2025}. Recall that this 20\% doping level corresponds to the regime where the superconductivity (SC) emerges so that it is plausible to expect for the role of the $s$ orbital in SC. 

Besides, it is worth noting that the spectra exhibits a larger broadening near the $X$ point. In fact, this is precisely the region where the $s$ orbitals indirectly influence the $d_{x^2-y^2}$-orbital dispersion discussed next.

\subsection{Suppressed $k_z$ dispersion of the $d_{x^2-y^2}$ orbital}

\begin{figure}[h!]
\centering
\psfig{figure=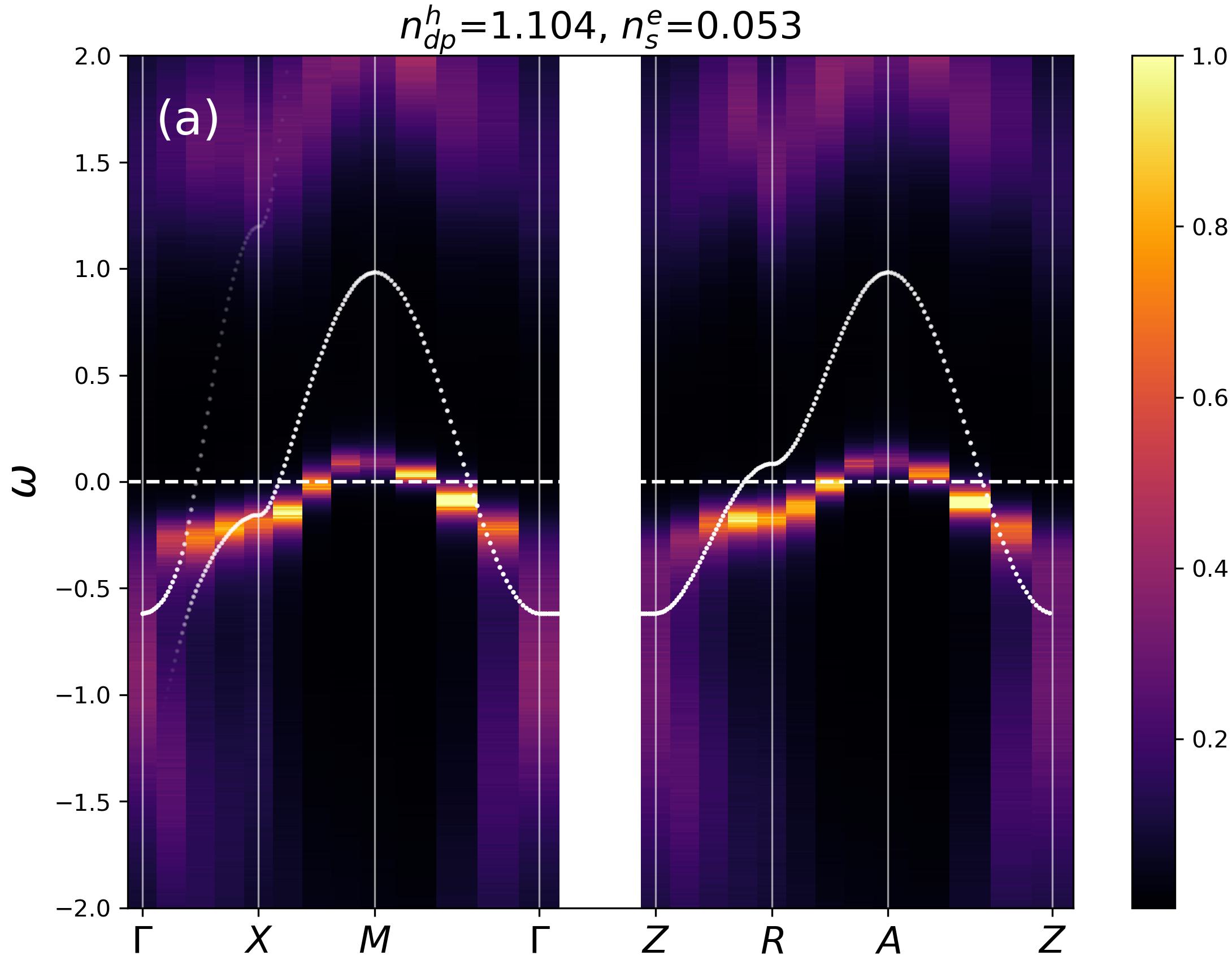, width=0.99\columnwidth}
\hfill
\psfig{figure=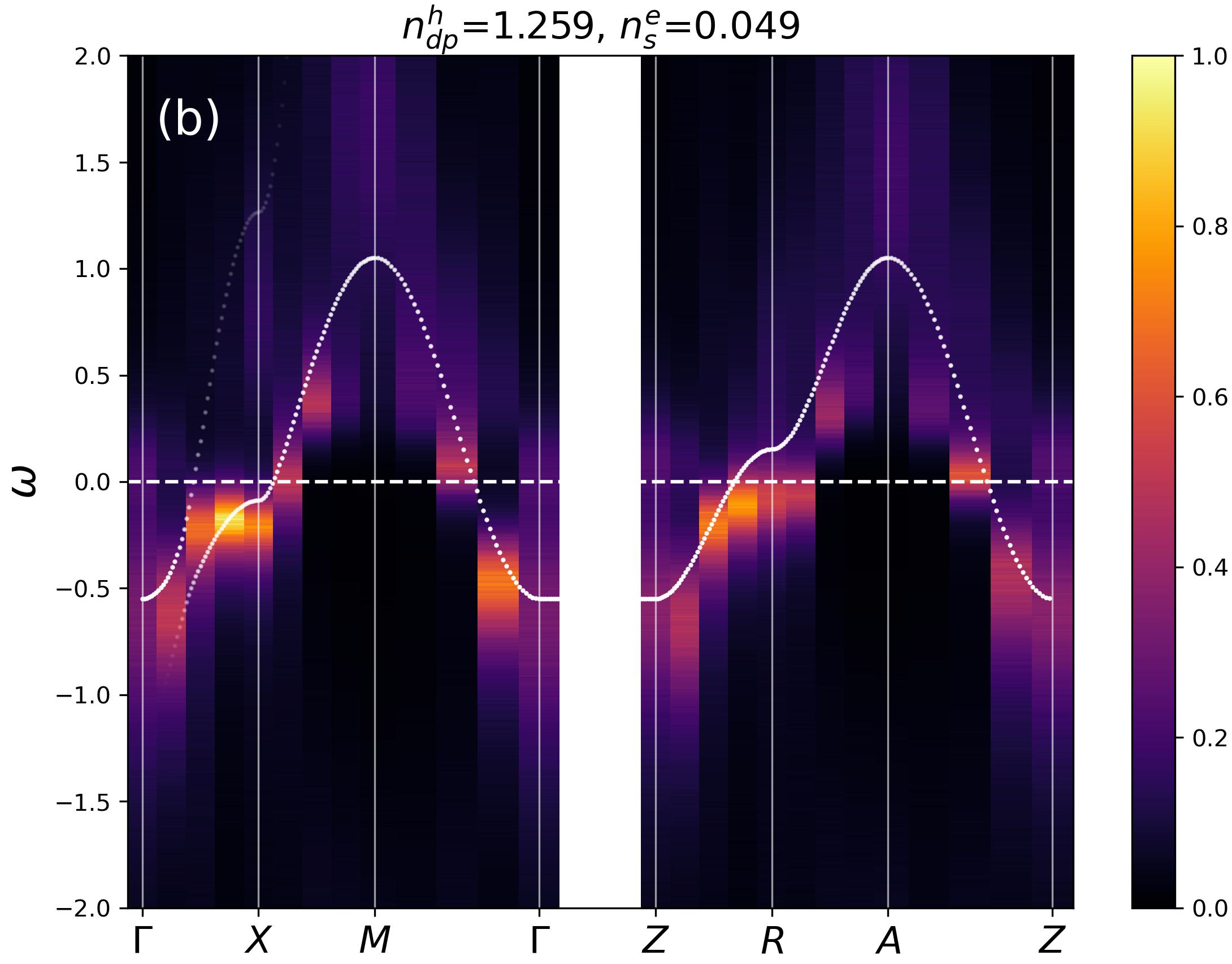, width=0.99\columnwidth}
\caption{Same as Fig.~\ref{Fig4}, but for the $d_{x^2-y^2}$ orbital.}
\label{Fig5}
\end{figure}

\begin{figure}[h!]
\centering
\psfig{figure=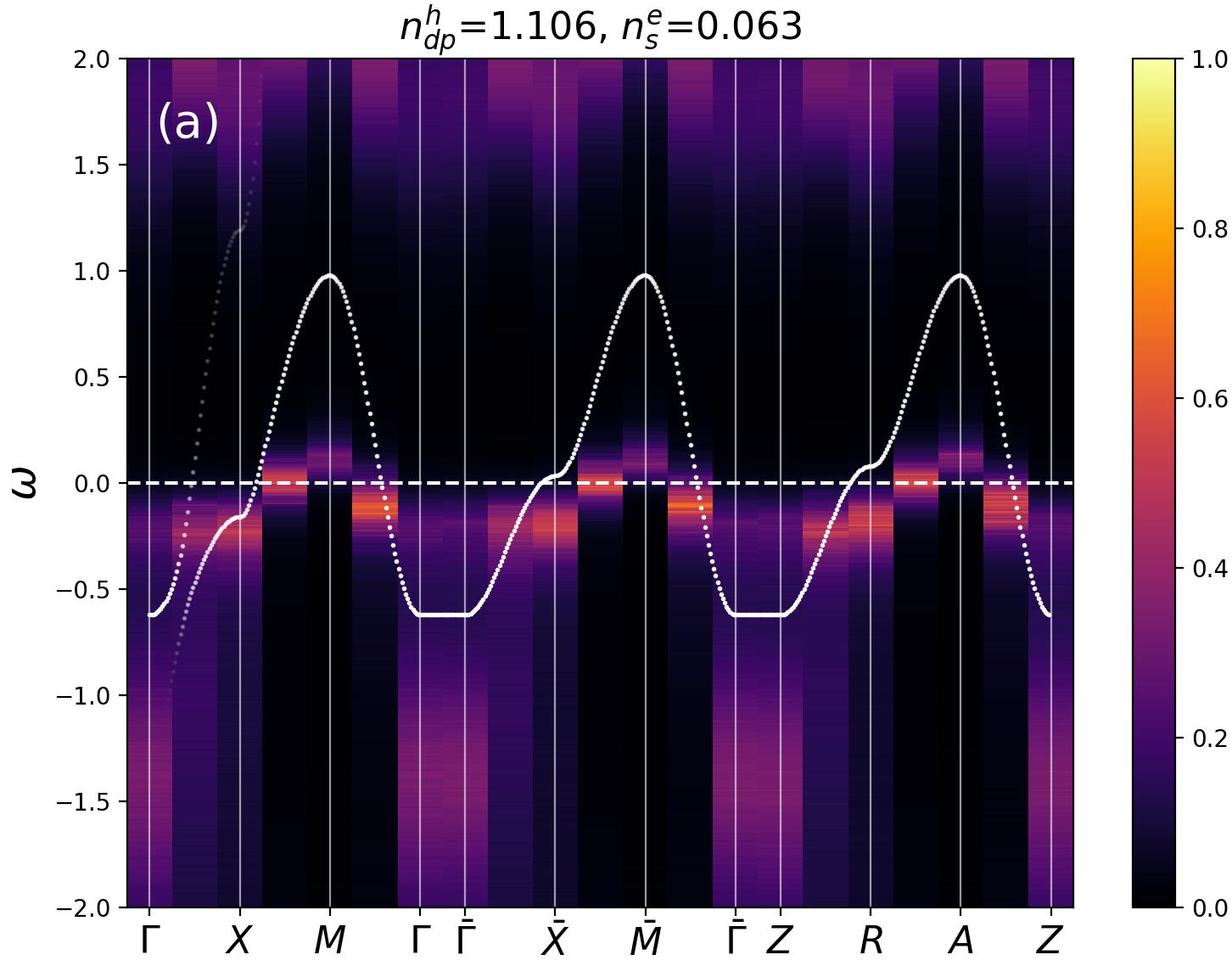, width=0.99\columnwidth}
\psfig{figure=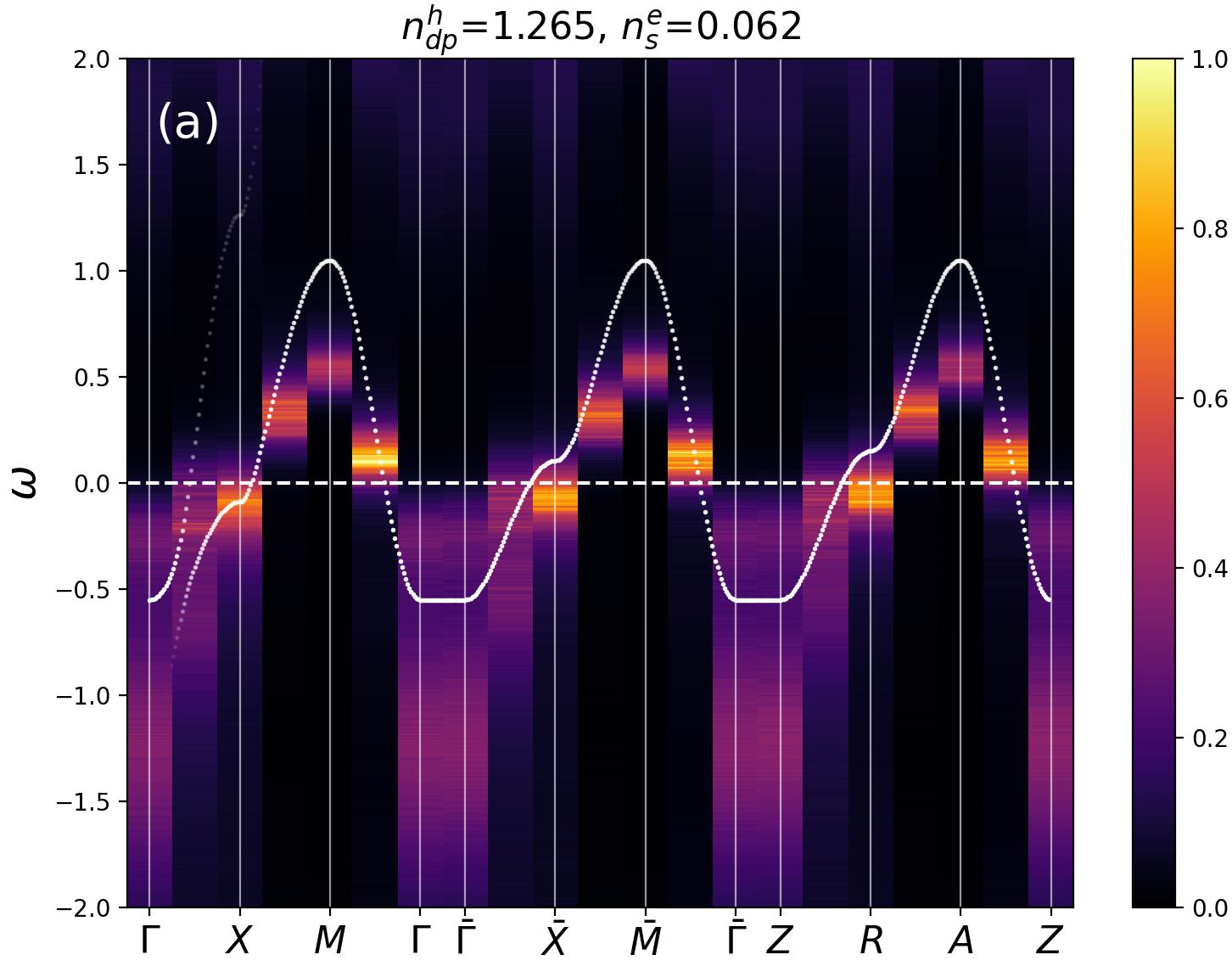, width=0.99\columnwidth}
\caption{Same as Fig.~\ref{Fig5}, but with a momentum resolution of $4\times4\times4$ and an inverse  temperature of $8.0$ due to a more severe sign problem.
 $\bar{\Gamma}$, $\bar{X}$, and $\bar{M}$ denote the corresponding high-symmetry points on the $k_z=\frac{\pi}{2}$ plane.
}
\label{Fig6}
\end{figure}

Another important consequence of strong electronic correlations is the strong suppression of the $k_z$ dispersion of the $d$-orbital band. In our $8\times8\times2$ DQMC simulations, the strong $U_{dd}=6.0$ eV renormalizes the bandwidth of $d$ orbital near the Fermi level and thereby leads to the low-energy behavior deviating from the DFT results. Fig.~\ref{Fig5}(a) corresponds to the parent IL nickelate compound, where the electronic dispersion along the high-symmetry paths in the $k_z=0, \pi$ planes do not display significantly different features except the region near $\Gamma$ and $Z$ points, although the hole pocket here is smaller than that observed in ARPES measurement~\cite{ding_cuprate-like_2024, li_observation_2025}. 

Even when the hole doping level reaches the experimentally relevant 20\%, as shown in Fig.~\ref{Fig5}(b), the dispersion still shows no indication of a Lifshitz transition~\cite{xia_three-dimensional_2025}. Meanwhile, most of the spectral weight of the UHB, originally located around 1.75 eV in panel (a), is transferred to the low-energy region in panel (b), whose spectra is closer to the non-interacting dispersion due to the doped charge carrier. 
As one of our major findings, Fig.~\ref{Fig5} vividly provides solid numerical evidence on the much weaker $k_z$ dependence of the dispersion than DFT calculations which also predicted the presence of Lifshitz transition~\cite{xia_three-dimensional_2025}.

In order to rule out the influence of finite-size effects arising from only two layers of $8\times8\times2$ lattice, we further carried out the simulation for a $4\times4\times4$ lattice with higher $k_z$ resolution at $\beta=8.0$. As shown in Fig.~\ref{Fig6}, the non-interacting dispersion (white dotted lines) of the $d_{x^2-y^2}$ orbital shows clear $k_z$ distinction between $X,\bar{X},R$ points; whereas the actual spectral function, regardless of undoped in panel (a) or doped in panel (b), still exhibits very weak $k_z$ dispersion along any high-symmetry path at fixed $k_z$. Unfortunately, due to the limitations associated with the QMC sign problem, we cannot explore even higher resolutions with a larger lattice at this stage. In other words, our large-scale many-body simulations do not see the strong $k_z$ dependence and the resulting Fermi surface change along $k_z$ direction~\cite{xia_three-dimensional_2025}.

\subsection{Enhanced antiferromagnetic spin correlation}

\begin{figure}
\centering
\begin{overpic}[width=0.49\textwidth]{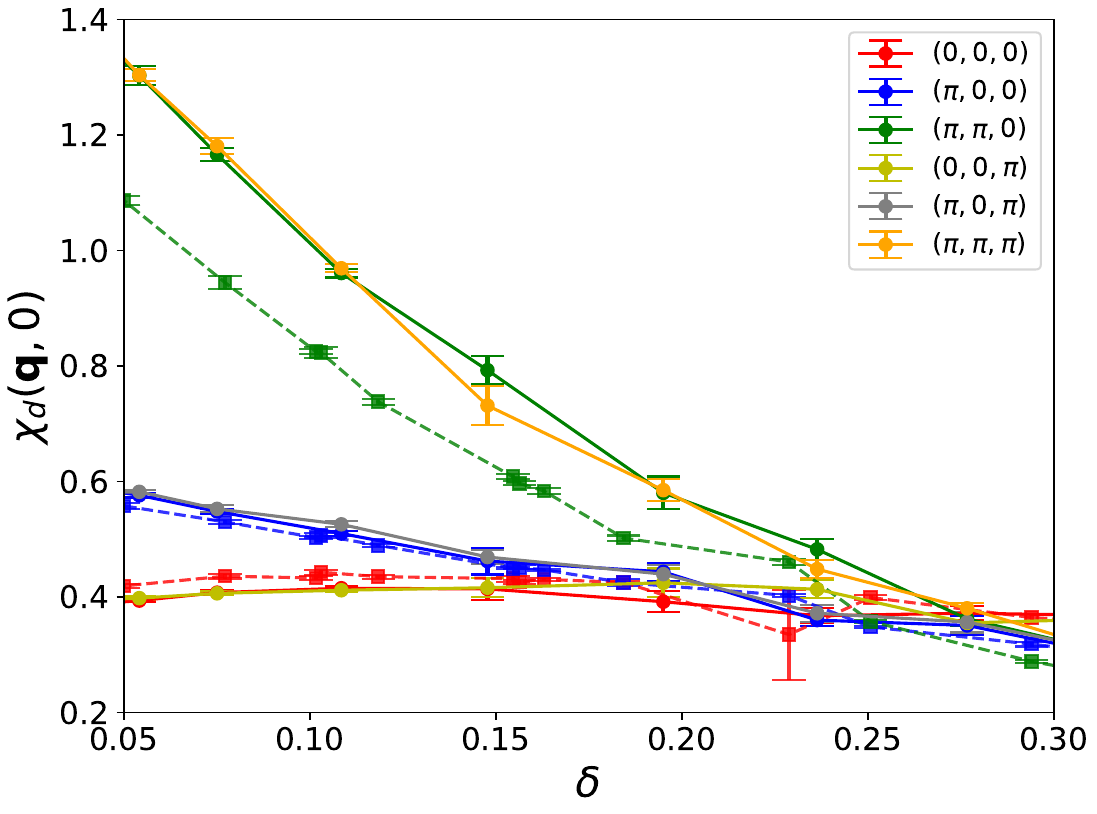}
    \put(35,48){
    \includegraphics[width=0.2\textwidth]{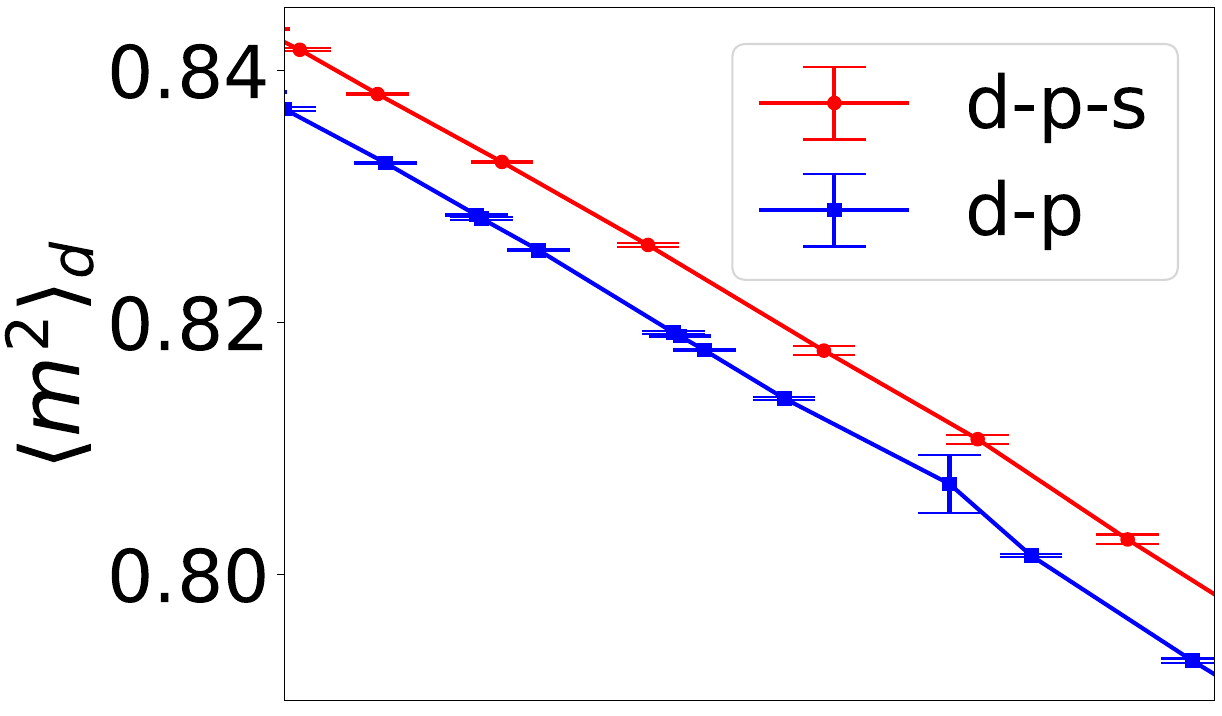}
    }
\end{overpic}
\caption{The static spin susceptibility $\chi_d(\mathbf{q},i\nu=0)$ of the $d_{x^2-y^2}$ orbital as a function of the hole doping $\delta=n_{dp}^h-1$ in the $d$-$p$ plane. The solid (dashed) lines are for $d$-$p$-$s$ ($d$-$p$) model. 
The inset shows the corresponding local moment $\langle m^2 \rangle_d$, with the same horizontal axis as the main panel.}
\label{Fig7}
\end{figure}

With the success of our many-body calculation to qualitatively reproduce the band structure revealed by ARPES experiments, which constitutes our major theme in the present work, to gather more information on the $d$-$p$-$s$ model, we continue to explore the momentum- and orbital-resolved zero-frequency spin susceptibility in a relatively small $4\times4\times2$ lattice, which is an important two-particle quantity for understanding the behavior of multi-orbital systems and is defined as 
\begin{align}
    \chi_{\alpha_1\alpha_2}(\mathbf{q},i\nu=0) = \sum_{\mathbf{l}} e^{i\mathbf{q}\cdot\mathbf{l}}\chi_{\alpha_1\alpha_2}(\mathbf{l},i\nu=0),
\end{align}
where 
\begin{align}
    \chi_{\alpha_1\alpha_2}(\mathbf{l},i\nu=0) = \frac{1}{N}\sum_\mathbf{i}\int_0^\beta \langle \hat{S}^z_{\alpha_1,\mathbf{l}+\mathbf{i}}(\tau) \hat{S}^z_{\alpha_2,\mathbf{i}}(0)\rangle d\tau
\end{align}
with $\alpha_{1,2}$ labeling the orbital, although here we focus on the case of $\alpha_{1,2}=d$ orbital.

Fig.~\ref{Fig7} presents the $\mathbf{q}$-resolved static spin susceptibility of $d_{x^2-y^2}$ orbital as a function of hole dopings $\delta$ covering a wide range $0.05 \le x \le 0.3$. The solid (dashed) lines correspond to the $d$-$p$-$s$ ($d$-$p$) model with the same set of parameters. 
One can see that the $d$-$p$-$s$ model exhibits apparently stronger in-plane $\mathbf{q}=(\pi,\pi,0)$ antiferromagnetic (AFM) spin-spin correlation than the $d$-$p$ model (green lines). Simultaneously, the ferromagnetic correlation at $\mathbf{q}=0$ is slightly smaller; while that at $\mathbf{q}=(\pi,0)$ remains nearly unchanged with the additional $s$ orbital. These observations indicate that the $s$ orbital leads to an enhancement of the antiferromagnetic fluctuations in the system. The inset displays the local moment of the $d_{x^2-y^2}$ orbital, which also displays the corresponding enhancement for all dopings. Put another way, the $s$ orbital indeed makes the $d_{x^2-y^2}$ orbital more localized and thereby promotes the AFM correlation between one another. 

Fig.~\ref{Fig7} also provides a direct comparison of the $q_z$ dependence of the static spin susceptibility in $d$-$p$-$s$ model. Note the nearly overlap between $\mathbf{q}=(0,0,0)$ (red) and $\mathbf{q}=(0,0,\pi)$ (yellow) solid curves as well as those between $\mathbf{q}=(\pi,0,0)$ (blue) and $\mathbf{q}=(\pi,0,\pi)$ (gray) curves.
More importantly, the cubic $\mathbf{q}=(\pi,\pi,\pi)$ AFM spin correlation (orange) is almost identical to the in-plane green solid line. 
All these features provide an alternative confirmation of the quasi-two-dimensional nature of the $d_{x^2-y^2}$ orbital in the $d$-$p$-$s$ model.

Last but not least, the enhancement of the AFM spin correlation has strong implication on the closely related superconducting instability, whose pairing glue is widely believed to originate from the AFM fluctuations in unconventional SC. We mention that our simulation indeed shows some signature (not shown) of enhanced $d$-wave pairing strength associated with $d_{x^2-y^2}$ orbital. Further exploration with more sophisticated methods are desired to decisively discriminate the exact role of the interstitial $s$ orbital now that it contributes to the Fermi surface even at relatively high doping as discussed for the band structure.




\section{Conclusion}
\label{sec:conclusion}
In this work, we investigated the electronic and magnetic properties of a $d$-$p$-$s$ model relevant to IL nickelates compared to the conventional $d$-$p$ model by employing large-scale DQMC simulation. 
Our results reproduce the essential electronic structure of both the parent compound and the 20\% hole doped IL system~\cite{ding_cuprate-like_2024, li_observation_2025}, which exhibit noticeable deviation from the original DFT dispersion without the renormalization from strong interaction. 

Specifically, the $d_{x^2-y^2}$ orbital exhibits strong localization and its $k_z$ dispersion is strongly suppressed, which is more compatible with the ARPES experimental findings than that predicted in DFT calculations. This strong quasi-2D nature persists even at 20\% doping without clear signature of the van Hove singularity and Lifshitz transition along the $k_z$ direction~\cite{xia_three-dimensional_2025}. The above weak $k_z$ dependence also manifests in the spin correlations of the $d_{x^2-y^2}$ orbital. 

Meanwhile, the interstitial $s$ orbital shows similar weak $k_z$ dependence and a smaller electron density than the DFT results. Instead its electron occupancy gradually decreases with hole doping but the $A$ electron pocket does not disappear even at 20\% hole doping. 

Moreover, compared with the $d$-$p$ model, the $d$-$p$-$s$ model shows enhanced AFM spin correlations on the $d_{x^2-y^2}$ orbital. Notably, recent nuclear magnetic resonance measurements~\cite{zhou_origin_2025} on IL nickelates have reported spin correlations stronger than those predicted by conventional theoretical estimate. Taken together, our results highlight both the strong electronic correlations associated with the $d_{x^2-y^2}$ orbital and the important role of the interstitial $s$ orbital in achieving better agreement with experimental observations for both the electronic structure and spin correlations of IL nickelates. 

Unfortunately, due to the limitation imposed by the fermionic sign problem, we are unable to access the regime of larger hole doping or lower temperatures. 
Future studies using alternative many-body approaches would therefore be valuable to further examine this phenomenon and to evaluate the corresponding superconducting pairing strength.

\section{Acknowledgments}
We acknowledge the support of the National Natural Science Foundation of China (Grant No.12174278, No.92477206) and State Key Laboratory of Surface Physics and Department of Physics in Fudan University (Grant No.KF2025\_12).
Mi Jiang also acknowledges the support of the China Scholarship Council (CSC) and the hospitality of Karsten Held at TU Wien.

\bibliography{main}

\end{document}